\documentclass[10pt]{article}
\usepackage{graphicx}
\usepackage{longtable}
\title{On
the second derivatives of periods  and braking indices in radio pulsars}

\author{
Malov I.\\
Pushchino Radioastronomical Observatory, 142290~Moscow Region, Pushchino, Russia}

\begin{document}

\maketitle

\begin{abstract}
The analysis of some braking mechanisms for neutron stars was carried out
to determine the sign of the second derivative of the pulsar period. This
quantity is the important parameter for calculations of  the braking
index n. It is shown that this derivative can be positive  and lead to
decreasing of n. It is necessary to correct the methods of calculations
of n used this moment because they are based as a rule on the suggestion
on the constancy of pulsar parameters (magnetic fields, angles between
some axes and so on). The estimations of corrections to braking indices
are obtained. It is shown that these corrections can be marked  for
pulsars with long periods and their small derivatives.

{\bf keywords}
braking mechanisms -- magnetic fields -- pulsars
\end{abstract}
\section{Introduction}

 One of the most important parameters characterizing evolution of pulsars
 is the so called braking index n describing the dependence of the angular
 rotation frequency on time:

\begin{equation}
 \frac{d\Omega}{dt}=-K \Omega^n,
\end{equation}

where  $K$ is a constant determined by the mechanism braking  the neutron
star.
 The quantity  $n$  can be calculated by the following expression:

\begin{equation}
n = \frac{\Omega d^2 \Omega / d t^2}{(d \Omega / d t)^2}
\label{eq2}
\end{equation}

  It is worth noting that (2) is correct for any braking mechanism if K is constant. Usually the following form

\begin{equation}
n = 2 - \frac{P d^2 P / d t^2}{(d P / d t)^2}
\label{eq3}
\end{equation}

is used  instead  (2).  Here  $P =\frac{2 \pi}{\Omega}$ is the rotation period. This period and its derivatives can be measured during long enough observations., and we can calculate on principle the braking index n and define the braking mechanism. However measurements of the second derivatives are complicated as a rule and to this moment they are determined precisely enough for several pulsars only. But there is the second difficulty in calculations of $n$ connected with the value of $K$. It is suggested usually that its dependences on magnetic induction, on inclination of magnetic moment to the rotation axis and other parameters do  not depend on time. If $K$ evolves  with the pulsar age  (2) and (3) demand some corrections. Blandford and Romani (1988) pointed out on the possible  importance of the dependence of $K$ on time many years ago. However the majority of authors believe  now that $K$ is constant.

Here we discuss the possible corrections for (2) and (3)  and signs of the second derivative of the rotation period.

\section{ Braking due to magneto-dipole radiation}

   The most popular braking mechanism in radio pulsar investigations is connected with the magneto-dipole radiation of a magnetized spherical neutron star. In this case the rate of losses of the rotation energy is equaled to the radiation power:

\begin{equation}
I \Omega \frac{d \Omega}{d t}  = -  \frac{2  \mu^2  \Omega^4 \sin^2 \beta}{3 c^3} ,
\label{eq4}
\end{equation}

where $I$  is the moment of inertia of the neutron star, $B_s$  the magnetic induction at the pole of the neutron star, $R_*$ the radius of the neutron star, $\beta$  the angle between the rotation axis and the magnetic dipole moment $\mu = \frac{B_s R_*^2}{2}$, $c$ the speed of light. It is suggested usually that  all quantities besides  $\Omega$   are constants and that $\beta = 90^{\circ}$. In this case (4)  leads to  $n = 3$. If     $I = 10^{45} \, g \, cm^2$ and $R_* = 10^{6} \, cm$, then we obtain the magnetic induction at the pole:

\begin{equation}
B_s = 6.4 \times 10^{19} \left( P \frac{dP}{dt} \right)^{1/2} \, G
\label{eq5}
\end{equation}

The known catalogs  (for example, Manchester et al., 2005) contain inductions at the magnetic equators. Their values are two times less than given by (5).

The progress of the magneto-dipole model can be connected with the work of Davis and Goldstein (1970) suggesting the exponential falling of  $\sin \beta$  with time:

\begin{equation}
\sin \beta = \exp (- t / \tau),
\label{eq6}
\end{equation}

Using the formula (2) we obtain from (6):

\begin{equation}
n = 3 + 2 \tan^2 \beta,
\label{eq7}
\end{equation}

This equality has been cited many times in much more late works. In (6)  $\tau$  is the characteristic time of decreasing of the angle $\beta$.  In the common case  the parameter $K$ does not constant and  (6)  means that $K$ as a function of $\beta$  must depend on time. The correct form for   $n$   is the following one  which has been given by  Blandford and Romani (1988):

\begin{equation}
n = \frac{\Omega d^2 \Omega / dt^2}{(\Omega / dt)^2}  - \frac{\Omega}{d \Omega / d t} \frac{d K /d t}{K}
\label{eq8}
\end{equation}

The model of braking due to magneto-dipole losses gives  in the case of  the constant magnetic field  and the evolution of the angle $\beta$  by the law  (6)

\begin{equation}
n = \frac{\Omega d^2 \Omega / dt^2}{(\Omega / dt)^2} + \frac{2 \Omega}{(d \Omega / dt) \tau}
\label{eq9}
\end{equation}

Philippov et al. (2014) gave the magneto-hydrodynamical model of the pulsar magnetosphere and showed that the evolution of the angle $\beta$  ran  slower than the exponential falling (6). For the dependence

\begin{equation}
\sin  \beta = \left( \frac{t}{\tau} \right)^{-1/2}
\label{eq10}
\end{equation}
we obtain

\begin{equation}
n = \frac{\Omega d^2 \Omega / dt^2}{(\Omega / dt)^2} + \frac{\sin^2 \beta}{\tau} \frac{\Omega}{(d \Omega / dt)}
\label{eq11}
\end{equation}

Using the period and its derivative  instead of $\Omega$  and $\frac{d\Omega}{dt}$  gives instead of (9) and (11) l the following expressions:

\begin{equation}
n = 2 - \frac{P d^2 P / dt^2}{(P / dt)^2} - \frac{2 P}{\tau d P / d t}
\label{eq12}
\end{equation}

and

\begin{equation}
n = 2 - \frac{P d^2 P / dt^2}{(d P / dt)^2} + \frac{P\sin^2 \beta}{\tau d P / d t }
\label{eq13}
\end{equation}

for the laws of the evolutions  (6) and (10), correspondingly.

Let us estimate the correction to the value  $n = 2.51$, calculated  using (3) for the Crab pulsar‚ B0531+21  (Lyne et al.,1993).  From the catalog of Manchester et al. (2005) we have $P = 33 \, msec$ and  $\frac{dP}{dt} = 4.2 \times10^{-13}$, and from the work of Loginov et al. (2016)   $\tau = 1.42 \times 10^6 \, years$. For these values of  the parameters  we obtain
$\Delta n = - 3.5 \times 10^{-3}$, i.e. for this pulsar the correction is inessential.  However for objects with the large characteristic age

\begin{equation}
\tau_c = \frac{P}{2 d P / d t}
\label{eq14}
\end{equation}

such a correction can be noticeable

As follows from (3) and (8) the value of n depends strongly on the sign of the second derivative of the period and on the value of $\frac{dK}{dt}$.

For the magneto-dipole mechanism the sign of $\frac{d^2P}{dt^2}$ coincides with the sign of the following polinomial:

\begin{equation}
\sin \beta \frac{d B}{d t} + B \cos \beta \frac{d \beta}{d t} -A B^3 \frac{\sin^3 \beta}{2 P^2},
\label{eq15}
\end{equation}

where

\begin{equation}
A =   \frac{8 \pi^2 R_*^6}{3 I c^3}
\label{eq16}
\end{equation}

We will omit the index  $s$  of the quantity  $B$ meaning that we  will deal with the magnetic induction at the surface. It is evident from (15) that for constant or falling with time values of  $B$ and  $\beta$  the derivative $\frac{d^2P}{dt^2} < 0$   and $n > 2$. Since the angle  $\beta$  decreases in this model the positive second derivative is possible for the increasing magnetic field only.

This moment there are no reliable data showing the decay of pulsar magnetic fields. On the other hand there are mechanisms of generation of magnetic fields during the pulsar evolution (see, for example, Blandford et al., 1983 and Sedrakyan and Movsisyan, 1986). Therefore the suggestion on the increasing field is not absurd. Suggesting similar to Philippov et al. (2014) that the decreasing of the angle  $\beta$  is very slow , omitting the second term  in (15), and putting $\sin \beta = \frac{1}{2}$, we obtain:

\begin{equation}
\frac{d B}{d t}  \ge  \frac{A B^3}{4 P^2}
\label{eq17}
\end{equation}

If the period $P$ grows linearly the solution of this equation  leads to the following inequality:

\begin{equation}
\frac{1}{B_0^2}- \frac{1}{B^2}  \ge  \frac{A}{2 d P / d t}  \left(\frac{1}{P_0} - \frac{1}{P} \right)
\label{eq18}
\end{equation}
\vspace{\baselineskip}

 Here index 0 means the values of parameters taken in the initial moment of time. To estimate the necessary growth rate of magnetic field  we put  $P_0 = 0.1 \, sec, \, P = 1 \,sec, \, \frac{dP}{dt} = 10^{-15}, \, B_0 = 10^{12} \, G, \, t = 10^6 \, years$. For these values of parameters the equality takes place  if  $B = 1.56 \times 10^{12} \, G$, i.e., .this growth is rather slow. Indeed for the exponential growth:

\begin{equation}
B = B_0 \exp \left( \frac{t}{\tau_B} \right) ,
\label{eq19}
\end{equation}

we have  $\tau_B \approx 2$  billions years.

The further observations are necessary to confirm or refute this effect.

Let us consider other models describing the slow down of the neutron star.

\section{Current losses}

In this model braking of the neutron star  connect  with currents on the surface  and their interaction with  its magnetic field.  This process leads to the evolution equation (Beskin et al., 1983):

\begin{equation}
I \Omega \frac{d \Omega}{d t}  = -  \frac{b i_0 B^2 R_*^6  \Omega^4 \cos \beta}{c^3} ,
\label{eq20}
\end{equation}

where  $b$  is the numerical coefficient equaled to  0.33 - 0.48 when the angle  $\beta$ changes from  $0^{\circ}q
$ to $90^{\circ}$, $i_0$ is dimensionless  longitudinal current  depending on   $\beta$ also. We have in this case instead of (15):

\begin{equation}
B \cos \beta \frac{d A_1}{d \beta} \frac{d \beta}{d t} + 2 A_1 \cos \beta \frac{d B}{d t} - A_1 B \sin \beta \frac{d \beta}{d t} - \frac{A_1^2 B^3 \cos2 \beta}{P^2},
\label{eq21}
\end{equation}

where

\begin{equation}
A_1 ( \beta) = \frac{4 \pi^2 R_*^6}{I c^3} b i_0
\label{eq22}
\end{equation}

Neglecting as earlier the dependence of the angle   $\beta$  on time we conclude that  the positive second derivative  is possible for $\frac{dB}{dt} > 0$ only. In this case

\begin{equation}
\frac{d B}{d t}  \ge \frac{A_1 B^3 \cos \beta}{2 P^2}
\label{eq23}
\end{equation}

Carrying out  calculations as for the case of the magneto-dipole braking we obtain the magnetic induction  $B = 1.25 \times 10^{12} \, G$  after 1 billion years. Hence in the current model it is necessary the slow secular growth of magnetic field  to achieve $\frac{d^2P}{dt^2} > 0$.

\section{Disk model}

Michel and Dessler (1981) have discussed the possibility  of the explanation of  pulsar peculiarities  suggesting the existence of a relic disk near the neutron star. Matter of this disk determines the structure of the pulsar magnetosphere and its braking. The corresponding equation of such a braking can be written in the following form:

\begin{equation}
I \Omega \frac{d \Omega}{d t}  = -  \frac{ \pi  B^2 R_*^6  \Omega^3}{ 3 G M} ,
\label{eq24}
\end{equation}

where   $G$ is the gravitational constant,  $M$ the mass of the neutron star. It follows from this equation:

\begin{equation}
\frac{d P}{d t} = \frac{2  \pi^2  R_*^6}{ 3 I G M}  B^2 = A_2 B^2 ,
\label{eq25}
\end{equation}
\begin{equation}
\frac{d^2 P}{d t^2} = 2 A_2 B \frac{d B}{d t}
\label{eq26}
\end{equation}

Thus in this model also the second derivative can be positive for $\frac{dB}{dt} > 0$ only.

\section{Current losses in the magnetosphere}

Electric fields and currents in the magnetosphere can lead to losses of energy (de Jager and Net, 1988). These losses can be described by the following equation:

\begin{equation}
I \Omega \frac{d \Omega}{d t}  = - \frac{k  B^2 R_*^5 \Omega^2}{c^2} ,
\label{eq27}
\end{equation}

where $k$ is a constant coefficient (less than 1). From (27) we have

\begin{equation}
\frac{d P}{d t} = \frac{k  R_*^5}{I c^2} P B^2  = A_3 P B^2
\label{eq28}
\end{equation}

and

\begin{equation}
\frac{d^2 P}{d t^2} =  A_3 \left( \frac{d P}{d t} B^2+ 2 P B \frac{d B}{d t} \right)
\label{eq29}
\end{equation}

This case differs from the previous models by the possibility of the positive second derivative not only  for the growing magnetic field but for the falling with time as well. In the last case the following inequality must be fulfilled:

\begin{equation}
\left| \frac{d B}{d t} \right| < \frac{A_3 B^3}{2} = \frac{k  B^3 R_*^5}{2 I c^2} ,
\label{eq30}
\end{equation}

For the used values of parameters this means that $\left| \frac{dB}{dt} \right|  < 0.56  \, G/sec$.

\section{Processes in the neutron star}

The circular motion of neutrons in the neutron star can lead to emission of neutrino-antineutrino pairs and to the dipole radiation (Huang et al., 1982). In this case energy of neutrons is passes to super-fluid vortexes and the neutron star is braking by the law:

\begin{equation}
I \Omega \frac{d \Omega}{d t}  = - \frac{11 \gamma^4 m_n^2 R_p^2  \Omega}{6 h^2 c^2  \langle n^* \rangle} \overline{\Delta^2 B^2   {n^*}^3} ,
\label{eq31}
\end{equation}

where $\Delta$ is the energy gap connected with the Cuper's pairs , $n_*$ a circulation quantum number of vortex, $R_p$ radius of  the super-fluid region,  $\gamma$  the neutron gyromagnetic ratio, $m_n$ its mass, h  the Plank's constant, the bar denotes the average for all the vortex lines. The equality (31) leads to the equation::

\begin{equation}
\frac{d P}{d t} = \frac{11 \gamma^4 m_n^2 R_p^2  P^2}{12 \pi h^2 c^2\langle n^* \rangle I}  \overline{\Delta^2 B^2   {n^*}^3}   =  A_4  B^2 P^2
\label{eq32}
\end{equation}

It is suggested  that the mean magnetic field inside the star is equal to the field at the surface. Taking as in the work of Huang et al. (1982) $\Delta =2.35 \, MeV, \, n_* = 10, \, R_p = 0.1 R_*$,  we have (Deng et al., 1987)

\begin{equation}
A_4 = 5 A
\label{eq33}
\end{equation}

It follows from (32) that
\begin{equation}
\frac{d^2 P}{d t^2} =  2 B P^2 A_4 \left(\frac{d B}{d t} + A_4 P B^3 \right)
\label{eq34}
\end{equation}

and to obtain the positive second derivative we must suggest either the growth of magnetic field  or its falling  with the rate

\begin{equation}
\left| \frac{d B}{d t} \right| < A_4 B^3 P = \frac{40  \pi^2 R_*^6 P B^3}{ 3 I c^3}
\label{eq35}
\end{equation}

For the used parameters this gives  $\left  | \frac{dB}{dt} \right  | < 4.87 \times 10^{-3}$ G/sec.

\section{Pulsar wind}

Particles escaping from the magnetosphere carry away an angular momentum. As a result there is the braking of the neutron star with the rate  (Harding et al.,1999)

\begin{equation}
I \Omega \frac{d \Omega}{d t}  = - \frac{L_p^{1/2} B R_*^3 \Omega^2}{ (6 c^3 )^{1/2}} ,
\label{eq36}
\end{equation}

where  $L_p$  is the power of the pulsar wind. The equality (36) gives

\begin{equation}
\frac{d P}{d t} = A_5 B P
\label{eq37}
\end{equation}

Here
\begin{equation}
A_5 = \frac{L_p^{1/2}  R_*^3}{ I   (6 c^3 )^{1/2}}
\label{eq38}
\end{equation}

and the value of the second derivative is determined by the following equality:

\begin{equation}
\frac{d^2 P}{d t^2} =  A_5 P \left(\frac{d B}{d t} + A_5  B^2 \right)
\label{eq39}
\end{equation}

This quantity is positive if $\frac{dB}{dt} > 0$. For $\frac{dB}{dt} < 0$, it is necessary to fulfill the following  condition:

\begin{equation}
\left| \frac{d B}{d t} \right| < \frac{L_p^{1/2} B^2 R_*^3}{I   ( 6 c^3 )^{1/2}}
\label{eq40}
\end{equation}

Taking  $L_p = 10^{33} \, erg/sec$, we obtain   $\left| \frac{dB}{dt} \right|  < 2.5 \times 10^{-3} \, G/sec$. This corresponds to the decay time of order of 10 billions years. Hence in the model of the pulsar wind both signs of the second derivative are possible.

\section{Propeller regime}

Sometimes an accretion from a debris disk on a neutron star can play a certain role in a braking of pulsars. In this case the so called propeller regime can be realized. In such a case we have (Illarionov and Sunyaev, 1975):

\begin{equation}
I \Omega \frac{d \Omega}{d t}  = - \frac{G M_*   d M / d t}{r_{eq}} ,
\label{eq41}
\end{equation}

where $M_*$ is  the mass of the neutron star, $\frac{dM}{dt}$ the rate of accretion,

\begin{equation}
r_{eq} =  \left( \frac{G M_*}{\Omega^2} \right)^{1/3}     -
\label{eq42}
\end{equation}

the distance where the rotation velocity is equal to the Kepler's velocity. The equation (41) can be transformed  to the following form:

\begin{equation}
\frac{d P}{d t} = A_6 P^{7/3}
\label{eq43}
\end{equation}

Here

\begin{equation}
A_6 = \frac{d M}{d t} \frac{(G M_*  / 4  \pi^2 )^{2/3}}{I}
\label{eq44}
\end{equation}

It follows from (43) that

\begin{equation}
\frac{d^2 P}{d t^2} = \frac{7}{3} A_6 P^{4/3} ,
\label{eq45}
\end{equation}

i.e. the pulsar rotation is braking during all time of its evolution with the increasing rate. It is worth noting that  the braking index is negative $(n= -1/3)$ in this regime.

\section{Discussion and conclusions}

Table 1 contains estimates of the braking index for 9 pulsars (Ho, 2015) calculated using the formula (2).

We have given the corresponding estimate for the Crab pulsar in the beginning of our paper. For the rest objects we have used  the formula (12) taking   $\tau  = 1.42$ billion years and obtained values of corrections  $\Delta n$, given in the last column  of the table. We can see that these corrections are small for all 9 pulsars. However we must point out once more that values of  $\Delta n$   can be noticeable for pulsars with long periods and small derivatives of the period.  It is follows also from the table  that the second derivative  must be positive for the pulsars  B0833-45, J1734-3333 and J1833-1034. For  B0531+21, B0540-69, J1119-6127, B1509-58 and J1846-0258 this derivative is negative.  In the case of the pulsar J0537-6910 we must expect  the influence of a debris disk. It is very important to search for such a disk around  J0537-6910.  We can not use the formula (7) for all pulsars from the table .

The value of the second derivative depends strongly on pulsar parameters. We will give one estimate only in the frame of the pulsar wind model. Omitting the term $\frac{dB}{dt}$ in (39) we obtain:

\begin{equation}
\frac{d^2 P}{d t^2} = A_5^2 P B^2 = \frac{L_p R_*^6 B^2 P}{6 I^2 c^3}
\label{eq46}
\end{equation}

Taking  $P = 1 \, sec, \, L_p= 10^{33} \, erg/sec, \, R_* = 10^6 \, cm, \, B = 10^{12} \, G, \, I = 10^{45} \, g \, cm^2$  we obtain $\frac{d^2P}{dt^2} = 6 \times 10^{-30} s^{-1}$.  Such derivatives we can expect in the precise timing measurements.

The choice of the braking mechanism remains the extremely important problem for the understanding of many  processes running in pulsars and  the determining of the ways of their evolution. As follows from our analysis new more precise estimates of the second derivatives are necessary. They will give the possibility to advance in the choice of the braking mechanism  and conclude on the changes with time  some pulsar parameters, in particular,  magnetic fields and the angles between the rotation and magnetic axes.

Hobbs et al. (2004) carried out the giant work on compilation of timing data  for more than 300 pulsars. They gave values of the second derivatives . However these values did not characterize the basic mechanisms of braking but were caused  by noises of different nature. Indeed there are no mechanisms giving values of n of order of  tenths or even thousands and both signs.  Unfortunately their data are not useful for the choice of braking mechanisms for individual pulsars. This moment only values from the table 1 can be used for this aim.

   There are   works where sone kinds of oscillations are postulated to explain large values of n (see, for example, Birykov et. al., 2012, Xie and Zhang, 2014).  They used  a number of suggestions and worked out  the so called toy models with many parameters. In any case they did not help to choise the  main braking mechanism.

This moment we can conclude that for pulsars with the measured second derivative corrections to the braking index are small and we can use formulas (2) and (3).

\section*{Acknowledgements}
This work has been carried out with the financial support of Basic Research Program of the Presidium of the Russian Academy of Sciences "Transitional and Explosive Processes in Astrophysics (P-41)". The author thanks L.B.Potapova  for the help with  the preparation of the manuscript.

\begin{table}[p]
\caption{ }
\begin{tabular}{|c|c|c|c|c|}
\hline
Pulsar & $P$ (sec) & $(dP/dt)_{-13}$ & $n$ & $- \Delta n$\\
\hline
    B0531+21&        0.033&         4.21&      2.51(1)&   0.0035\\
   J0537-6910&        0.016&         0.52&    -1.5(1)&   0.014\\
   B0540-69&        0.050&        4.79&     2.087(7)&   0.0047\\
   B0833-45&        0.089&        1.25&     1.4(2)&   0.032\\
  J1119-6127&        0.408&        40.20&    2.684(2)&   0.0045\\
   B1509-58&        0.151&        15.31&    2.832(3)&   0.0015\\
  J1734-3333&      1.169&        22.79&    0.9(2)&   0.023\\
  J1833-1034&        0.062&        2.02&    1.8569(6)&   0.0137\\
  J1846-0258&       0.327&      71.07&    2.65(1)&   0.0021\\
\hline
\end{tabular}
\end{table}

\end{document}